# Positive Trust Balance for Self-Driving Car Deployment

Philip Koopman[1,2], Michael Wagner[1]

[1] Edge Case Research, Pittsburgh PA 15201, USA
[2] Carnegie Mellon University, Pittsburgh PA 15213, USA
koopman@cmu.edu, mwagner@ecr.ai

**Abstract.** The crucial decision about when self-driving cars are ready to deploy is likely to be made with insufficient lagging metric data to provide high confidence in an acceptable safety outcome. A Positive Trust Balance approach can help with making a responsible deployment decision despite this uncertainty. With this approach, a reasonable initial expectation of safety is based on a combination of a practicable amount of testing, engineering rigor, safety culture, and a strong commitment to use post-deployment operational feedback to further reduce uncertainty. This can enable faster deployment than would be required by more traditional safety approaches by reducing the confidence necessary at time of deployment in exchange for a more stringent requirement for Safety Performance Indicator (SPI) field feedback in the context of a strong safety culture.

**Keywords:** Self-driving cars, autonomous vehicles, system safety, deployment.

## 1    Introduction

At some point, developers must make a decision that it is time to deploy a Self-Driving Car (SDC) design. Ultimately, all the testing and safety engineering efforts come down to a binary go/no-go decision: is the vehicle ready to operate on public roads with sufficiently safe outcomes? This paper proposes an approach to living with the uncertainty that will be inherent in making this decision.

At the time of deployment, the SDC design team must be able to show that ***expected operational safety will be acceptable***. However, with the still-maturing state of SDC technology, it is likely that there will be significant uncertainty surrounding any such expectation. Moreover, due to practical limits on simulation and human-supervised testing, the scale of operations required to establish confidence that the error bars on the expectation also fall within acceptable safety outcomes will be too large to achieve in any way other than collecting data from actual at-scale deployment. (We use the term "confidence" in a general mathematical sense of a confidence interval and the like, and not in the sense of the strength of an individual's subjective belief.)

This results in a cyclic dependency: the only way to resolve uncertainty is to deploy, but deployment cannot be justified based on a high confidence expectation of acceptable safety due to excessive uncertainty. We propose a framework to ensure a responsible deployment decision based on Positive Trust Balance (PTB), involving a combination of validation, engineering rigor, post-deployment feedback, and safety culture.

A key observation is that traditional safety engineering is based on the premise that the first deployed unit is safe enough to be part of an at-scale production run and that – potential product recalls notwithstanding – no further changes will need to be made after deployment to achieve the desired lifecycle safety target. Such a claim might be unsupportable via practicable validation efforts for salient aspects of SDC technology such as the use of machine learning based systems. However, rather than saying that it is hopeless to assure safety at deployment, we instead propose an approach in which the *expected* level of safety is shown to be acceptable, but there is a non-negligible potential for higher than desired operational risk exposure due to *uncertainty*. This might still result in net tolerable risk if post-deployment feedback can be relied upon to aggressively reduce uncertainty over the system lifecycle.

In a related work, [Kalra17] assumed that safety was perfectly characterized when deploying, and explored the implications of uncertainty of in potential safety improvements after deployment. In contrast, we deal with uncertainty of the expected safety at the time of deployment.

## 2      Current Approaches to Deployment Decisions

### 2.1      Positive Risk Balance

A decision to deploy should include a decision as to whether the system is expected to be "safe enough." While there is no universally accepted criterion, a common approach is that SDCs should be at least as safe as a human-driven car. If the SDC is safer than a human, it is said to have Positive Risk Balance (PRB) [DiFabio17].

Typically, PRB is stated in terms of the SDC having a lower fatality rate than otherwise comparable human-driven vehicles. A more nuanced view should also encompass major injury rates, minor injury rates, and perhaps property damage events. Setting a credible risk target is not trivial, and should take into account comparable loss event rates for the target Operational Design Domain (ODD), risk distribution profile (e.g., whether vulnerable road users are put at increased risk), and numerous other considerations (e.g., according to Section 16 of [4600]) rather than simply using a generic national-level statistic for fatality rates. Other safety postures are possible, such as comparison only with unimpaired human drivers, or even a goal of zero at-fault loss events.

For present purposes we assume some reasonable definition of PRB is the goal. That assumption having been made, the question is how to decide whether PRB will be achieved before deploying.

### 2.2      Driver Test

A commonly proposed deployment criterion is passing some sort of simulated and/or real-world driving road test, potentially drawing upon elements of human-equivalent road tests and a scenario catalog (e.g., [Cerf18]).

An operational test is essential to confirm the validity of simulation and analysis. However, any predetermined test will struggle to assess analogues to some real-world human driving skills such as the ability to handle novel unstructured situations and "common sense" contextual interpretation. While such attributes are not stressed in human driver test procedures, traditional driver licensing addresses them indirectly via minimum age, supervised instruction, and brief observation of behavior by a human driver test official. Such an approach is deemed sufficient based upon significant experience with human abilities and cognitive development plus mandatory insurance. There is no analogous testing-only approach to evaluate judgment maturity for SDC technology for licensing and insurance risk evaluation purposes.

### 2.3 Testing Metrics

Administering a comprehensive SDC driving test requires significant resources, so a common alternative approach has been to use metrics that reflect on-road testing experience. Example metrics are number of miles driven, automation disengagements [Banerjee18], and crashes per mile. (It is important to note that physical testing on public roads presents potentially significant risk that must be mitigated [Koopman19].) More sophisticated approaches combine simulation results with actual road miles.

While it is difficult to justify a deployment decision for a vehicle that lacks substantial real-world testing, large-scale testing campaigns don't necessarily ensure safety. Potential threats to validity for a road testing safety campaign must also be addressed, including changes to underlying vehicle software mid-campaign, driving only "easy" miles, and in general driving in conditions that do not cover risky portions of the ODD.

Even if road testing addresses all experimental concerns, the sheer number of miles required (likely billions of miles [Kalra17]) is infeasible to conduct using physical road tests before deployment.

### 2.4 SOTIF Approaches

The need for billions of miles of testing can, in principle, be reduced by identifying scenarios based on real-world operation and ensuring that the vehicle performs properly for all scenarios possible within the ODD. A methodical iterative improvement approach to this is used for Safety of the Intended Function (SOTIF) based workflows [ISO21448][SAFAD19]. The general idea is to iteratively identify and mitigate so-called triggering events that expose requirements gaps or other functional insufficiencies, resulting in ever-expanding scenario catalog. It is generally assumed that the deployment decision will be based on having high confidence that risk is not unreasonable. However, if a heavy tail distribution of triggering events is present, it might be that it is economically infeasible to discover enough of the "unknown hazardous" scenarios to achieve such high confidence, even if all available evidence based on mitigation of known hazards supports a conclusion of being safe enough to deploy.

The 2019 revision of ISO/PAS 21448 describing a SOTIF approach includes both iterative improvement during development and a newly added field monitoring section. Both activities are essential to achieve practical safety, and are aligned with the approach described in this paper. However, it is beneficial to also define a more explicit relationship between initial deployment safety and the role of field feedback in making the initial deployment decision, especially in the presence of substantial uncertainty.

## 3    The Positive Trust Balance Approach

It seems likely that a combination of the approaches discussed will be required, including analysis, simulation, testing, scenario catalogs, and iterative improvement. But, even doing all these things together is likely to result in an unacceptable level of uncertainty about whether a PRB has actually been achieved before deployment. The proposed four-prong Positive Trust Balance (PTB) approach addresses this uncertainty.

The issue is that design validation techniques are based on leading metrics that *predict* a PRB, rather than lagging metrics that confirm a PRB has been achieved in practice [Rand18]. And, even if pre-deployment on-road PRB lagging metrics could be gathered (at great expense), it would likely be unaffordable to repeat full-scale on-road evidence collection before deploying each periodic software update.

A prime motivation for quick SDC deployment is to mitigate losses attributed to human drivers. However, a key deployment risk is uncertainty as to whether PRB will actually be achieved, or if instead losses will be worse than with human drivers due to premature deployment. Fundamental sources of uncertainty (and therefore lack of confidence) that complicate understanding of advanced SDC technology risk include:
- Lack of a human-comprehensible design for tracing tests back to design intent as is typically done with a V-model development cycle.
- Still-maturing best practices for developing machine learning-based systems.
- Addressing the problem of how an SDC can know that it is operating (or about to operate) outside its ODD when it encounters an unforeseen edge case.

Resolving this uncertainty requires more than yet another testing or simulation tool. Rather, it requires a fundamental re-thinking of the goal of having conclusive evidence that a safety target has been reached before deploying a system.

We suggest considering a PTB approach that involves an ***initial practicable, evidence-based expectation that a safety target will be achieved, and then uses operational data to improve the confidence of that expectation over time.*** It is essential to have a robust safety culture for this approach to be viable.

### 3.1    Pre-Deployment Validation

Pre-deployment validation, including analysis, simulation, and testing, should be carried out to the maximum extent reasonably practicable. That having been said, it has long been known that brute force testing is impracticable for establishing the safety of high criticality systems (e.g., [Littlewood93]). Therefore, we should accept that a testing-only approach will leave substantial uncertainty as to safety.

Consider a hypothetical validation strategy:
— 10,000 million mile simulation campaign
— 100 million miles of collected road data used to feed those simulations
— 10 million miles of actual road testing used to validate simulation results

Even if the road testing proceeds with no incidents whatsoever, there will be uncertainty as to how well the SDC can handle infrequent events that arrive too seldom to be thoroughly characterized by the road data and road test campaign. As a simple illustration, a substantial fraction of rare events that happen once every 100 million miles won't have been seen at all in that data collection, let along be seen in road testing. But some of those same comparatively rare events will in aggregate occur more frequently than every 100 million miles during deployment. That will in turn potentially invalidate any PRB validation claim of a 100 million mile or longer average fatal mishap arrival rate. This is in addition to residual concerns about simulation accuracy that will further increase uncertainty.

Consider a developer claim that road testing and simulation has been done to the limits of economic practicality. The outcome of such a successful limited testing campaign will be that developers believe the system is safe – *as far as they know*. But, that knowledge of safety will have low confidence if it is not based on more data collection and testing than the expected fatality interarrival rate, which is generally impractical. Put another way, the developer will have an expectation of reasonable safety, but the error bars will be too big for comfort due to economic constraints.

*Something will need to be done about the error bars.*

## 3.2     Engineering Rigor

Functional safety standards require not only testing, but also evidence that a sufficiently rigorous engineering process has been applied. In particular, the conventional software features of the SDC should be designed according to established safety standards, such as ISO 26262 [ISO26262] for functional safety.

While engineering practices for some aspects of SDC technology such as machine learning are still evolving, there are known bad practices to be avoided, and the use of best practices should be confirmed. For example, training machine learning-based systems on safety validation data is clearly undesirable, but in practice corners might inadvertently be cut or data management mistakes might be made. Best practices are not a panacea, but it should be established that they have been followed. Moreover, it is difficult to justify that questionable engineering practices will result in a safe SDC. Conformance to SDC-specific safety standards and guidelines can help with this (e.g., [4600][ISO21448][SaFAD19]).

## 3.3     Feedback and Continuous Improvement

Even though validation and engineering rigor have been used to attain an acceptable *expected* safety outcome, confidence will likely be low enough that an unacceptable outcome might yet occur (e.g., the mean risk outcome is acceptable, but figurative error bars also encompass unacceptable outcomes). To address this concern, feedback from actual operations should be used to improve confidence as well as fix any problems.

A traditional recall approach that waits for unambiguous trends in mishaps and only then issues reactive fixes is insufficient here. No developer should deploy an SDC suspected or known to be unacceptably safe and then wait for multiple fatal crashes to accumulate before taking corrective action. *The degree of safety uncertainty inherent to novel SDC technology incurs an obligation to proactively monitor operations and to respond to all incidents, including near misses.*

A specific approach to feedback is the use of Safety Performance Indicators (SPIs) as defined in ANSI/UL 4600 [4600]. SPIs are operational metrics that cover not only lagging indicators, but also leading indicators. SPIs are tied directly to an SDC's safety case. Near miss reporting [Williamsen13] is an essential part of this approach. The aviation industry uses SPIs to improve after near misses and process breakdowns without waiting for actual loss events to drive lessons learned [ICA009].

While it is always possible to get unlucky, in general if near misses are much more common than loss events, monitoring and correcting root causes of near misses can increase confidence faster than losses occur. Safe failure fraction SPIs are useful, such as the ratio of road testing near misses to operator-prevented mishaps.

This approach also encompasses the notion of small scale SDC pilot deployments. These still require a decision that the SDC is expected to be safe enough to deploy at pilot scale followed by feedback and improvement to build confidence over time.

### 3.4    Safety Culture

As with the design of any safety critical system, a robust safety culture [NASA15] is essential to providing acceptably safe SDCs. Some particularly important issues are:
- Avoiding setting unreasonably low initial quality and validation goals based on an argument that post-deployment updates will fix bugs. Such an argument is not aligned with the PTB approach.
- Using the lack of maturity in accepted practices in some areas (e.g., still evolving best practices for safe machine learning) as an excuse for not following well known best practices for more traditional aspects of the system, such as functional safety.

The need for a robust, transparent robust safety culture should not be news to anyone involved in safety critical system design. *With a PTB approach it is absolutely essential to have a strong safety culture* both to ensure good technical outcomes as well as enable public trust to be built over time.

## 4      Conclusions

A Positive Trust Balance approach to self-driving car deployment includes all of: testing, employing engineering rigor, using field feedback for continuous improvement, and building a transparent, robust safety culture. Rather than requiring likely unattainable conclusive proof that a risk target has been met on day one, instead the deployment decision is made based on practicable evidence collection that supports a reasonable expectation that risk will be sufficiently low. This must be coupled with a firm commitment to improve confidence in that expectation using post-deployment feedback. A robust and transparent safety culture is essential to ensure that the developer is committed to making acceptable safety decisions over the system lifecycle.

In the end, what will matter the most is whether stakeholders trust the safety of SDCs at least as much as they trust the safety of human-driven vehicles (i.e., Positive Trust Balance). We believe that strong tool support for a PTB approach embodied in a safety case will be crucial for combining these elements in practice and developing stakeholder trust.